\documentclass[manuscript,screen, authorversion,nonacm]{acmart}

\AtBeginDocument{%
  }

\setcopyright{acmlicensed}
\copyrightyear{2025}
\acmYear{2025}
\acmDOI{XXXXXXX.XXXXXXX}

\acmConference[Conference acronym 'XX]{Make sure to enter the correct conference title from your rights confirmation emai}{June 03--05,  2018}{Woodstock, NY}

\setcopyright{none}
\renewcommand\footnotetextcopyrightpermission[1]{}
\usepackage{dcolumn}
\newcolumntype{d}{D{.}{.}{2}}

\acmISBN{978-1-4503-XXXX-X/18/06}

\begin{document}



\title[Exploring Emotional Speech Commands as a Compound and Playful Modality]{Speech Command + Speech Emotion: Exploring Emotional Speech Commands as a Compound and Playful Modality}





\author{Ilhan Aslan}
\email{ilas@cs.aau.dk}
\orcid{0000-0002-4803-1290}
\affiliation{%
  \institution{Aalborg University}
  \city{Aalborg}
  \country{Denmark}
}

\author{Timothy Merritt}
\email{merritt@cs.aau.dk}
\orcid{0000-0002-7851-7339}
\affiliation{%
  \institution{Aalborg University}
  \city{Aalborg}
  \country{Denmark}
}

\author{Stine S. Johansen}
\email{stinesl@cs.aau.dk}
\orcid{0000-0002-8451-5444}
\affiliation{%
  \institution{Aalborg University}
  \city{Aalborg}
  \country{Denmark}
}

\author{Niels van Berkel}
\email{nielsvanberkel@cs.aau.dk}
\orcid{0000-0001-5106-7692}
\affiliation{%
  \institution{Aalborg University}
  \city{Aalborg}
  \country{Denmark}
}

\renewcommand{\shortauthors}{Aslan et al.}

\begin{teaserfigure}
\includegraphics[width=\linewidth]{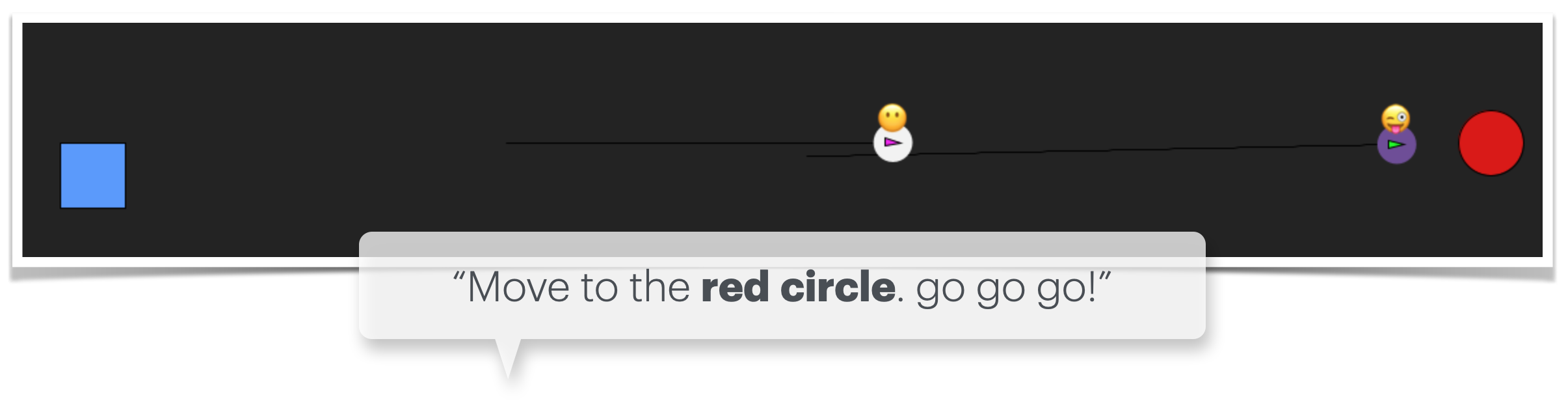}
\caption{Screenshot from our study showing the response to a speech emotion command, with one of two agents making use of speech emotions to modify how it implements the emotionally charged command to ``Move to the red circle, go go go!'' This agent displays a playful emoji and moves more quickly to the target, while the other agent displays a neutral emoji and moves more slowly to the target.}
\label{fig:teaser}
\end{teaserfigure}

\begin{abstract}
In an era of human-computer interaction with increasingly agentic AI systems capable of connecting with users conversationally, speech is an important modality for commanding agents. By recognizing and using speech emotions (i.e., how a command is spoken), we can provide agents with the ability to emotionally accentuate their responses and socially enrich users' perceptions and experiences. To explore the concept and impact of speech emotion commands on user perceptions, we realized a prototype and conducted a user study (\textit{N}~=~14) where speech commands are used to steer two vehicles in a minimalist and retro game style implementation. While both agents execute user commands, only one of the agents uses speech emotion information to adapt its execution behavior. 
We report on differences in how users perceived each agent, including significant differences in stimulation and dependability, outline implications for designing interactions with agents using emotional speech commands, and provide insights on how users consciously emote, which we describe as `voice acting'.
\end{abstract}






\maketitle

\section{Introduction}

The emergence of language technologies is enabling new and advanced voice interactions, driving visions of future computers as intelligent agents with expert language skills. Thus, a renaissance of speech interfaces and spoken language as a widespread modality to interact with computers in different forms seems likely.
While speech is recognized as an accessible interaction modality~\cite{corbett16} that allows users to command various technology, such as turning on lights in smart homes, choosing playlists in cars, and asking a mobile for assistance in everyday tasks, typical commands rely on predefined text patterns and make little use of humans' vocal range. This neutral and explicit command approach misses key aspects of typical human conversation that add richness and detail, such as emotional tone, which can impart a sense of urgency, add humor, or convey frustration.

AI's growing ability to perceive nuances and act with agency means agents can understand more than just explicit commands. Therefore, it is important to investigate how to utilize the broader spectrum of human vocal expression (e.g., tone, prosody, emotion) for richer human-agent communication. We propose to design embodied agents that can both recognize speech command and speech emotions (e.g., tone of voice) information to better connect with users, acknowledging the subjectivity in users' expressions, and rendering otherwise mundane experiences more memorable, reflective, and alive. There are both risks and opportunities of using speech emotions as a novel design material and both need to be explored and better understood. To this end, we realized a prototype to study the concept of emotional speech commands using a speech emotion recognition (SER) model. We treat speech emotion and speech command in speech as modalities that can be used by an agent for different purposes. 
With this prototype, we conducted a user study (\textit{N}~=~14), allowing participants to simultaneously instruct and steer two agents embodied as vehicles carrying an emoji back and forth between two targets (see Figure~\ref{fig:teaser}). While both agents implemented the participants' speech commands by identifying the navigation target, only one of them (the affective agent) implemented the concept of emotional speech command using both speech emotion and speech command as a compound modality to adapt its movement pattern and expression. 

Our results show significant differences in how participants experienced the affective agent vs the standard agent, including the affective agent' quality to stimulate, engage, and connect with users, which we argue can, depending on the design goal, be used to deliver customized affect transitions.
These can be beneficial for various domains including entertainment and education. We outline implications for the design of speech commands as an input modality.

\section{Background}

This paper relates mainly to research in multimodal interaction, affective computing, and speech-based human-agent interaction.

Speech as an interaction modality has been there from the early days of designing human-computer interactions and envisioning futures of how humans will interact with computers~\cite{lowerre1976harpy}. However, the limitations of using speech alone in an unimodal interaction setup has also been criticized, with fellow researchers arguing how co-located human interaction is naturally multimodal and that humans use other types of communication signals that contextualise spoken words~\cite{oviatt2015paradigm}.
Richard Bolt's seminal research~\cite{Bolt_put_that_there_1980} on combining speech and gestures by enabling an agent to contextually interpret the speech command ``Put That There''  has been a landmark for multimodal interaction research. Thereafter, multimodal interaction has been known for its potential to expand interaction spaces (e.g.,~\cite{Yoon13, Pen_Midair18}), and improve robustness and reduce uncertainty through fusing interaction modalities (e.g.,~\cite{mertes2024affecttoolbox, Lingenfelser11}). Consequently, Oviatt et al.~\cite{Oviatt01122000} have argued that multimodality expands computing to more challenging applications, a broader spectrum of users, and accommodate more adverse usage conditions, such as automotive~\cite{Sasalovici25} and tourism~\cite{aslan2005compass2008}.

Challenges in designing affective interaction have been also recognized early in HCI research~\cite{Boehner05, Boehner07} acknowledging a perspective of emotions as experience co-interpreted as they are made in interaction. Researchers have designed for a large body of affective experiences exploring, for example, bodily modalities to handle experiences of grief~\cite{Ravn24}, to support coping with loneliness~\cite{Passler24}, foster empathy~\cite{piHearts_2020}, or help playfully regulate emotions~\cite{Soler24}. 

With increasing progress in the field of machine learning, the capabilities and performance in computer audition are improving. In this context, human speech has become a rich source for recognizing various aspects of the user's context. It has been used to diagnose diseases, such as Alzheimer's disease, Parkinson's disease, depression, and COVID-19~\cite{Milling22}. The field of speech emotion recognition has benefited from progress in deep learning~\cite{Latif23}. This progress driven by deep learning has been crucial, since only ``now'' can we recognize emotional expressions in a person's voice at finer levels so that the exploration of speech emotions as an additional modality is made possible. 

Beyond other contextualising signals, human speech is in itself a source for multimodality as it involves, for example, linguistic information as well as mental and physical performance. Similar to how the choice of words in a language offers some degree of freedom for expressing a statement or a command, physical performance influences how the command sounds, including high-level features such as loudness and pitch describing voice infection. As social beings, humans have learned to use a range of expressions to communicate their intents and influence others.

We take inspiration from previous research~\cite{aslan2025speejis} augmenting voice messages with emotionally expressive visual cues, such as automatically adding emojis to voice messages. We replicate their use of  emoji classifications provided by Kutsuzawa et al.~\cite{kutsuzawa2022classification} and the same machine learning model~\cite{wagner2023dawn} for speech emotion recognition. In the present paper, the model accepts audio (in various lengths) as input and provides valence, arousal, and dominance values as output. The model is trained on audio input only,thus recognizes emotions based on acoustic features. However, as Wagner et al.~\cite{wagner2023dawn} describe, their model has emergent abilities to recognize emotional keywords. Overall, we deemed this model suitable to study the concept of speech emotion commands and enable agents to recognize ``how a command is spoken.''  To this end we rely on the abilities of the specific SER model, which is capable to map audio data containing speech to the well-known emotion dimension pleasure (valence), arousal, and dominance. Thus, an agent can for example extract a users' level of arousal from their speech command. However, in practice what we say and how we say it is often congruent, meaning that we tend to say nice words in a nice tone and bad words in a bad tone.  Consequently, even if the model does not understand all the words in a message, it should be able to recognize nuanced differences in commands that contain similar textual content, such as \emph{``STOP you dumb thing, go back to the blue!''} compared to \emph{``Please turn around and go to blue. Go, go, go!''}

Affective feedback to speech based user inputs can be provided in different modalities. For example, emojis are well known as affective symbols mainly in the context of messaging and augmenting messages with visual cues~\cite{kutsuzawa2022classification, aslan2025speejis}. Related work has also studied the influence of movement and animation~\cite{pretouch_proxemcis17, movetobemoved16} with researchers exploring, for example, the impact of a mobiles capable of moving or changing shape~\cite{Hemmert13,Pedersen14}. They argue that such feedback is associated with nature and living agents, such as animals; and that movement as feedback is generally associated with ``being alive''~\cite{Parkes08}. Non-verbal audio feedback is also an option to provide affective feedback, however, synthesizing such audio feedback while doable~\cite{Ritchel19} is compared to using existing emojis cumbersome. 

While we focus in this paper on speech as an input modality and the impact of non-verbal emotional feedback from the agent, it is still worth noting that speech can also be used as an output modality for affective interactions. This is especially true for a dialog setting compared to a setting where users command a machine. A survey on research on the impact of agent voice in voice-based human–agent interaction is provided by Seaborn et al.~\cite{Seaborn2022} encompassing research on voice agents, robots, and smart things including vehicles. Especially within human-robot interaction, speech has also been investigated as an output modality, demonstrating various factors that affect people's perceptions of the agent, including trustworthiness~\cite{torre2020if}, social presence~\cite{Lubold2016}, and reciprocal interactions~\cite{Ostrowski2021}. 

\section{Speech Command Prototype}
To study the experience of emotional speech commands as a playful and compound modality to interact with (intelligent) agents, we build a prototype inspired by the original ``Put that there'' demo~\cite{Bolt_put_that_there_1980}, in which a researcher demonstrates possible interactions combining speech commands with pointing gestures to create and move simple shapes, such as blue circles, red squares, and yellow triangles. 

\begin{figure}[ht!]
\includegraphics[width=\columnwidth]{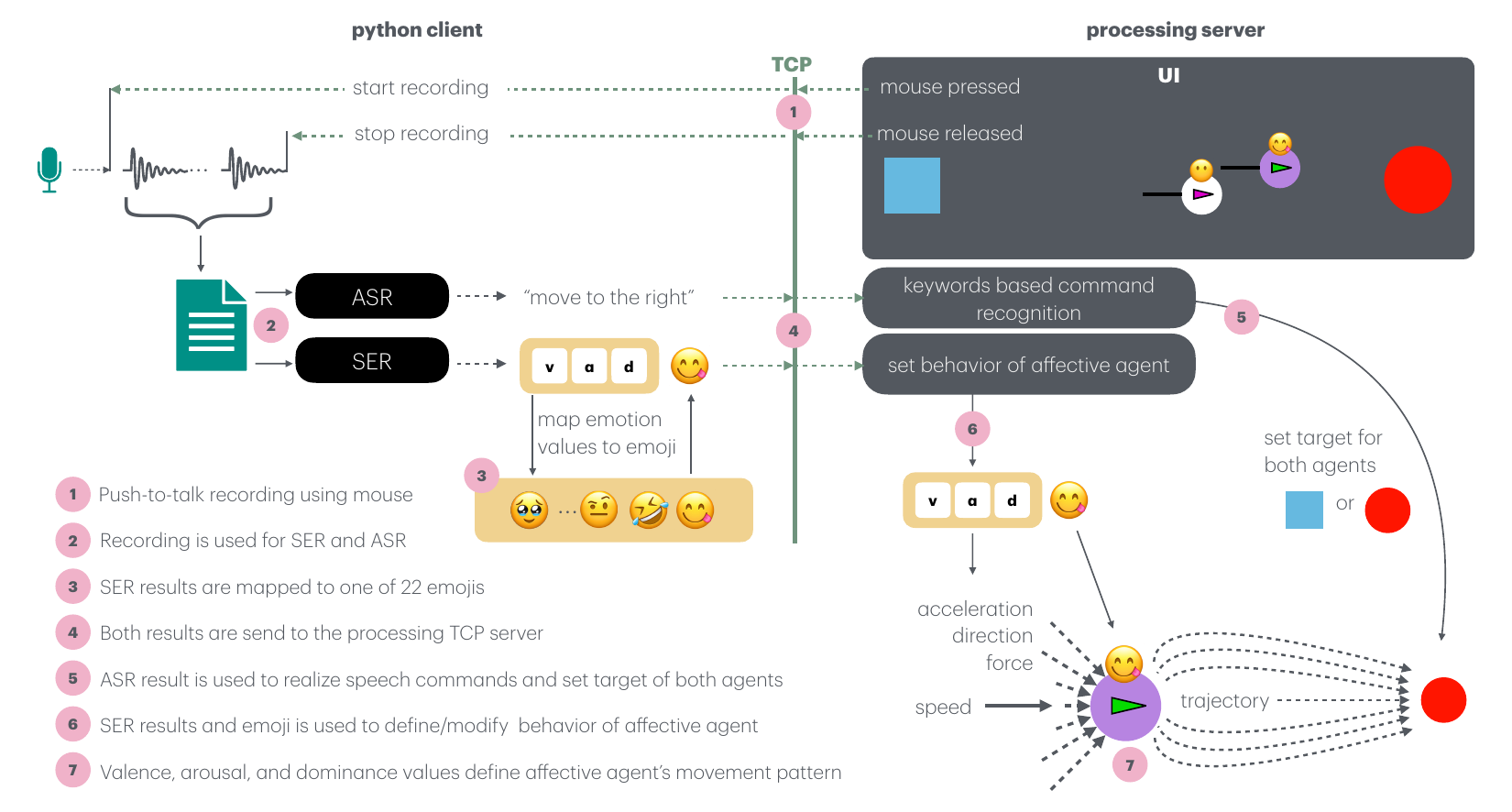}
\caption{Overview of the system architecture of the prototype.}
\label{fig:architecture}
\end{figure}

Figure~\ref{fig:architecture} provides an overview of our prototype's system architecture. The user interface is straightforward: we use a red circle and blue square on the left and right end of the application window as targets. The user can instruct two agents represented on the UI as `spaceships' to move to these targets by speaking commands such as ``move to the red circle'', ``move right''.
To recognize the commands we first apply automatic speech recognition (using the base Whisper model~\cite{radford23a}) and subsequently run keyword detection on the transcribed text. Here, we identify the words `red', `blue', `right', `left', `circle', and `square' to recognize the command and set the target. When the command is recognized, the spaceships moves to the target to comply with the users command. The standard agent uses the same speed and a straight path to move toward the target, while the affective agent also takes into consideration how the user spoke the command using a speech emotion recognition (SER) model extracting emotions (i.e., valence, arousal, and dominance values) from the audio. These emotion values are used to modify speed, force, and direction of initial acceleration vectors, which cause the agent to have a different movement pattern when implementing the command. The initial acceleration and force applied after receiving the command causes the agent to veer slightly upwards or downwards depending on valence and dominance in the speech, while the agent's force and speed are influenced by the arousal value. In addition, both agents are visually augmented with an emoji. The standard agent always is augmented with the neutral emoji, while for the affective agent we use the method described in previous research~\cite{aslan2025speejis} to map SER results in a voice message to one of 22 different emojis based on arousal and valence values.

\begin{figure*}[ht!]
\includegraphics[width=.8\textwidth]{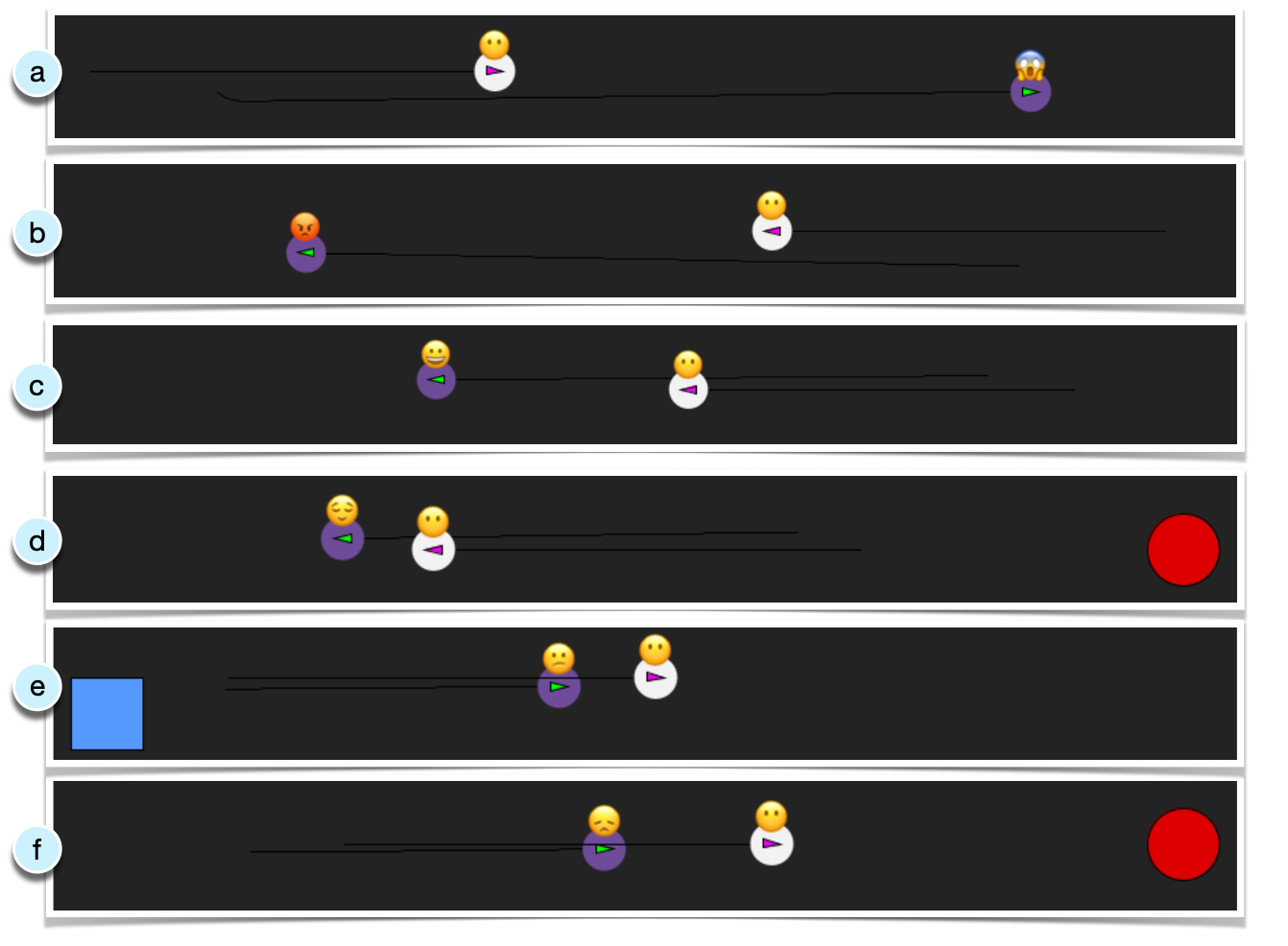}
\caption{Cropped screenshots showing examples of different behaviors of the agents for speech commands expressed in different ways. Lower position such as in a) and b) indicate lower valence, while higher speed indicates higher levels of (emotional) arousal in the speech command. c) is an example for high arousal high valence compared to f) which is lower arousal and lower valence. d) and e) are closer to a  a command given in a neutral tone. }
\label{fig:examples}
\end{figure*}

The system architecture consist of a TCP server application written in Processing and a TCP client application written in Python. For the speech commands we implemented a push-to-talk interaction principle, in which the user holds the mouse button pressed while talking. Releasing the mouse button signals the end of the speech command. While the mouse button is pressed the audio signal is captured with a microphone and the resulting audio data provided as input to both an ASR and a SER model. We treat the output of each of these model as separate input modalities and map the output of each for different purpose. The standard agent only makes use of the ASR output to recognize the speech command. The affective model additionally uses the SER output to modify implementation behavior (e.g., movement behavior and emoji).
To implement the `spaceships' and their movement we build on code provided by Shiffman's vehicle example written for Processing~\cite{shiffman2012nature}. Figure~\ref{fig:examples} shows examples of the agent behaviors across a range of valence and arousal, cropped out from screenshots while interacting with the prototype. 

\section{User Study}
While previous research has explored speech command interfaces extensively, there is a lack in research exploring the combination of speech command and speech emotion as a compound modality. Our assumption leading into the study was: an agent that implements the concept of emotional speech commands would be perceived differently from an agent that implements a standard speech command interface and  does not listen or adjust to speech emotions. Thus, the main research question we aimed to explore was: 

\emph{``What is the impact of an agent using both speech emotion and speech commands on user perception of the interaction with the agent and the agent itself?''}

We conduct a user study in which we use our experimental system (i.e., our prototype) to enable our participants to simultaneously interact with the two different agents. Both agents receive and respond to the same speech (emotion) commands, and thus, their reactions and infused experiences become directly comparable.  

\subsection{Participants and Procedure}
We recruited 14 participants (five female, nine male), four students and ten staff members from our university campus between the ages of 24-46 years (M=32). The study was conducted in a standard single-person office. We invited participants to take a seat in front of a desktop monitor and used a desktop condenser microphone to both record the speech commands and the posthoc interviews with the participants. At the beginning of the study, participants received a short introduction to what they were seeing on the (external) monitor, which appeared as a simple retro game (the app window with width 2500px and height 1300px nearly covers the whole screen). The app run on a Macbook Air (M4). We told them that they could command the two agents through speech. We explained that both agents would implement their commands but only one would also listen to how they spoke the commands and adapt its behavior while implementing the command. They were instructed that the agents only recognized key words, such as ``right'', ``red'', ``square'' embedded in their sentences to identify the target in their commands. We told participants that they could only send the agents back and forth between the two targets and that they did not have to wait until the the agents reached their target to state the next command. We informed participants that the affective agent would also listen to how the participants spoke to the agents to adapts its implementation and that if they wanted to cheer the agent up or be insulting, they could try that out.  Each participant was instructed to start by pressing the mouse and stating ``turn on the light''. The response to this command was that each of the agents displayed their lights (spheres around the vehicles). This first command was to ensure that participants understood how to perform speech commands based on push-to-talk. 

Participants were instructed to send the agents back and forth until they had understood and experienced the type of interaction and were ready to rate their user experiences. We asked participants to provide us verbal confirmation when they were ready. 

Once the participants were ready to provide their ratings, they were provided with the UEQ questionnaire (without the items measuring novelty, since it was clear that the new technique was highly novel). We also asked them to fill out the English version of the `perceived intelligence' from the Godspeed questionnaire~\cite{Bartneck2023}. 
Finally, we asked them to rate their agreement on a 5-point Likert scale (from strongly disagree to strongly agree) for the following opinions:

\begin{itemize}
\item I felt in control.
\item The agent collaborated with me.
\item The agent was engaging.
\item The agent understood me.
\item The agent reacted emotionally intelligent.
\item I connected with the agent.
\end{itemize}

These opinion ratings where used to structure the posthoc interviews asking participants to reflect on their understanding of the question and reasons for their ratings.  

\subsection{Results}
We first present the results for the UEQ and Godspeed questionnaires, including results of the statistical tests. Then, we move to illustrating the opinion ratings and the qualitative analysis of the posthoc interviews.

\begin{figure}[ht!]
\includegraphics[width=0.8\columnwidth]{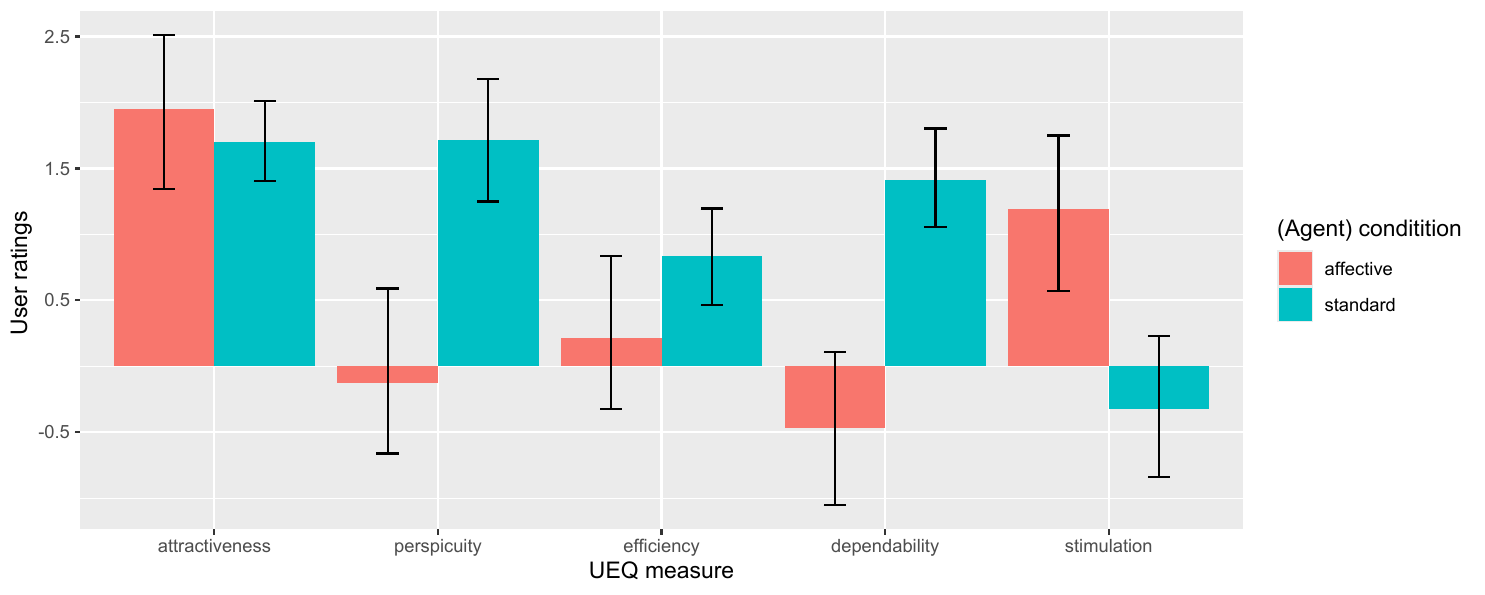}
\caption{Overview of UEQ scores, measuring the user experience of both agents. Error bars denote 95\% confidence intervals.}
\label{fig:UEQ}
\end{figure}

\begin{table*}[t]
\caption{Overview of UEQ results (p-values have been adjusted with Holm-Bonferroni correction).}
\label{tbl:ueqscores}
\centering
\begin{tabular}{l dd dd r dl}
\toprule
 & \multicolumn{2}{c}{Affective agent} & \multicolumn{2}{c}{Standard agent} \\
 \cmidrule(r){2-3} \cmidrule(l){4-5}
Dimension & \multicolumn{1}{r}{Mean} & \multicolumn{1}{r}{SD} & \multicolumn{1}{r}{Mean} & \multicolumn{1}{r}{SD} & t-statistic & \multicolumn{2}{c}{p-value} \\
\midrule
Attractiveness & 1.95 & 1.1 & 1.7 & 0.6 & t(13) = -0.71 & 1.00 &  \\
Perspicuity & -0.1 & 1.2 &  1.7 & 0.9 & t(13) = 4.36 & 0.004& $\ast$$\ast$\\
Efficiency & 0.21 & 1.1 &  0.8 & 0.7 & t(13) = 1.58 & 0.68 & \\
Dependability & -0.46 & 1.1 &  1.4 & 0.8 & t(13) = 4.58 & 0.003 & $\ast$$\ast$ \\
Stimulation & 1.2 & 1.2 & -0.3 & 1.1 & t(13) =- 3.13 & 0.039 & $\ast$ \\
\bottomrule
\multicolumn{8}{l}{\textit{Note:} $\ast\, p < 0.05$; $\ast$$\ast\, p < 0.01$} \\
\end{tabular}
\end{table*}

Figure~\ref{fig:UEQ} presents the UEQ ratings for both agents (affective vs standard) and Table~\ref{tbl:ueqscores} results of the statistical tests. The agent condition had a main effect on the dimensions `Perspicuity' and `Dependability' which relate to usability and `Stimulation' which relates to the user experience. While participants rated the affective agent as less usable they also rated the interaction with the affective agent as more stimulating.

\begin{figure}[ht!]
\includegraphics[width=0.8\columnwidth]{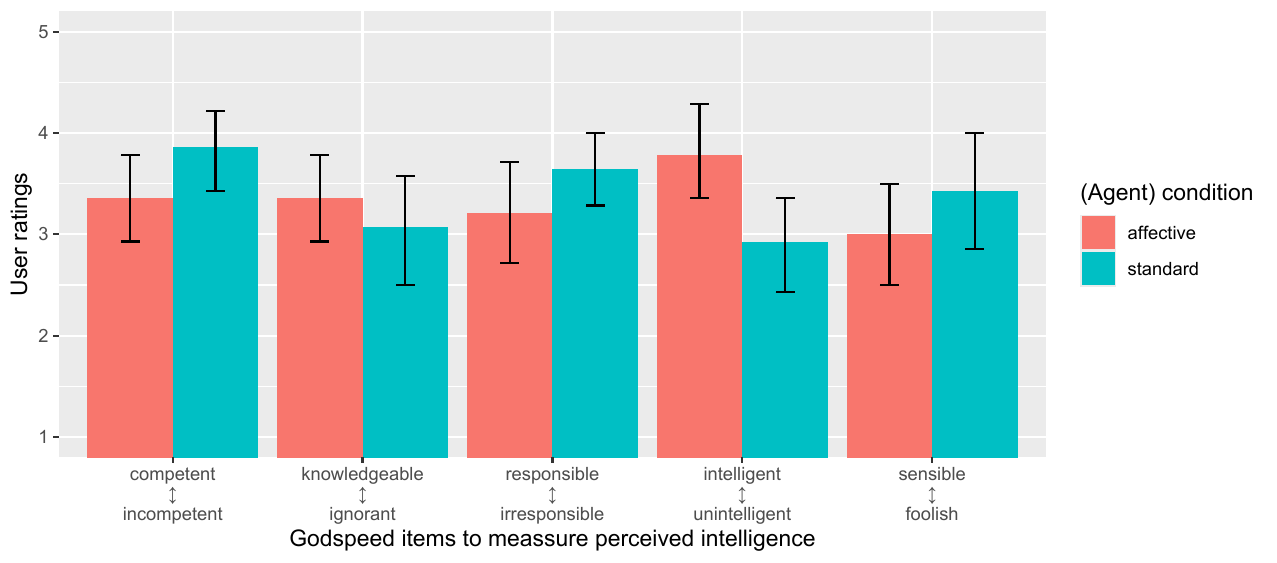}
\caption{Overview of ratings for the godspeed questionnaire items of the perceived intelligence module. Error bars denote 95\% confidence intervals. }
\label{fig:godspeed}
\end{figure}

Considering the participants' ratings for the perceived intelligence module of the Godspeed questionnaire, we found no significant effect of agent modality $t=0.15$ and $p=0.88$, with similar mean ratings for the affective agent ($M=3.34$ $SD=0.78$) and the standard agent ($M=3.38$ $SD=0.57$) based on a 5-point Likert scale. The ratings can be interpreted as participants attributing some intelligence to the agents but, overall, they attributed low intelligence for both of these agents.
Figure~\ref{fig:godspeed} presents the mean values for each item from the perceived intelligence module, contributing to the overall value. The individual item ratings provide some insights into how the participants' perception was different, e.g., the affective agent was rated as more foolish but not ignorant, and less responsible but more intelligent. Consequently, the qualitative results shed light into the different forms of ``intelligence'' that users associated with the agents.

\begin{figure}[ht!]
\includegraphics[width=0.8\columnwidth]{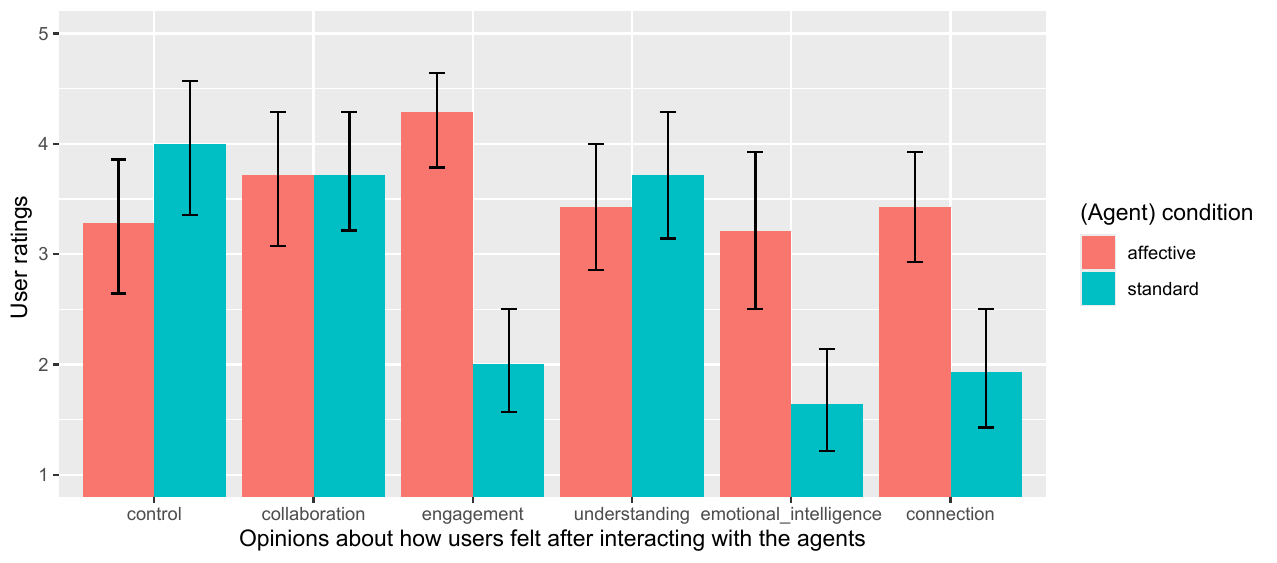}
\caption{Overview of mean opinion scores, provided by participant to describe how they felt about the agents and interacting with them. Error bars denote 95\% confidence intervals.}
\label{fig:MOS}
\end{figure}

In Figure~\ref{fig:MOS}, we present participants' mean opinion scores. The opinion ratings were used as a foundation to interview participants about their interpretations of the opinion statements and reasoning behind their ratings. To this end, we followed the basics of thematic analysis~\cite{Braun01012006} and protocol analysis~\cite{ericsson1980verbal}, which included the creation of mind maps for each opinion and merging them iteratively to a few key descriptors and connections, summarizing the results with respect to the subjective statements that we introduced to structure the analysis of the interviews.

In the following, we present the results of the thematic analysis, considering each of the topics introduced with each statement.

\textbf{I felt in control.} The main argument for participants who provided the same opinion ratings for both agents was that both agents had implemented their commands and did what they were told to. Participants who provided different ratings for how they felt in control explained that their feelings for the affective agent was different due to difference in expectations and predictability (e.g., \emph{``I think the standard one, there was never any doubt about what it would do''} (P10) and \emph{``I felt more in control of the consistent one''}) (P9), effort required to control (e.g., \emph{``I had to force myself to speak in a certain way to get control''} (P2) and \emph{``You get this additional layer that is, that I am not used to interacting through voice commands. This layer of voice acting''}) (P9), and attribution of having its own agency \emph{``It felt like it (the affective agent) had more agency of it own than the standard agent''} (P3).

\textbf{The agent collaborated with me.} 
There was some overlap of participants' ratings considering arguments they also made for how they felt control. A few participants argued that they did not feel a difference in collaboration as their understanding of collaboration entails a shared task. For those who rated feelings of collaboration for the agent differently, control over the speed of the agent was brought up as an aspect indicating more willingness of the agent to collaborate. Similarly, concepts relating to being social and having agency were brought up to explain why some participants ratings were higher (e.g., \emph{``Even though I said I felt more in control with the standard one I think it (the affective one) felt a bit more human or smart or something''}) (P10)
or lower \emph{``LOL, that one (the affective one) was more like its own, there was something it was doing beyond the command''} (P11) and \emph{``There was also a social thing there you know. Don´t defy me!''} (P9) for how they felt the agent collaborated with them.

\textbf{The agent was engaging.} Participants felt that the affective agent was (much) more engaging. Their explanations for rating that way included fun and playfulness of the interaction (e.g., \emph{``It was fun to play around with it (the affective one), I didn´t even look at the standard one''} (P8) and feeling more for the affective one (\emph{``I felt much more for this (affective agent) than for this (standard agent)''} (P5) and a sense of reciprocity sometimes reflecting a feeling as if the agent had expectations from the user (e.g., \emph{``when you see that it gets scared or it gets "Oh I better go fast" it feels like it is somehow communicating even though you know it is just a machine but the communication is there and that is quite nice. It feels like you are engaging with it and it is engaging with you back.''} (P6) and \emph{``it (affective one) got me to try to get it go faster to speed up, so it got me engaged in that sense.''}) (P2). 

\textbf{The agent understood me.} Participants explained that their lower ratings for the affective agent related to multiple interpretations for the agent's responses (e.g., \emph{``this one (affective one) understood it  too but I don´t think it always change the speed, or when I said something it didn´t always match like what I wanted it to do'')} (P10) and realizing a conflict between how and what is being said (e.g. \emph{``If it had the connection between my command and how I said it. For example, if I said go slow in a happy voice I would want it to still look happy while going slower'')} (P5) with the agent only listening to how something is said. Participants reflected on the idea that understanding is more bidirectional with the affective agent, where the agent understands the user but the user also wants to understand the agent's reactions. One participant described the agent's reactions and the emotional mismatch between user and agent as \emph{``So sometimes I felt like "OK you are afraid now?, it wasn't that bad what I was saying", so a bit dramatic, also the speed the idea of "OK slow down, I am not your mum"''} (P7).  

\textbf{The agent reacted with emotional intelligence.} Participants explained their higher ratings for the affective agent by providing examples of interactions. For example, one participant tried out fake laughing during the command and then had to really laugh based on the affective agent's reactions, stating \emph{``When I laugh it reacted and went faster but it was also because I was laughing naturally. It was kind of funny how it reacted to the fake laugh like it was shocked, which I would say is emotional intelligence''} (P3) and another participant explained \emph{``when I tried to sound angry it kind of felt like it was rushing "OK I need to finish this quickly becuase she is angry", but in that sense that was the funny part. There was a bit of comedy, where I felt like it was a collaboration and I was angry and the agent was working for me in a sense and I could understand why it would feel like it needed to rush''} (P4). There was also an understanding that the ``intelligence'' was different (e.g., \emph{``I might not know why it is reacting the way it is reacting but it seems to me that it has some emotional intelligence but it doesn't have 100\% emotional intelligence. I also sometimes did not understood if it was trying to mimic me, reflect back what I was doing, or it is reacting to my voice''} (P7). Participants associated an entertainment value with the affective agent which was also brought up as an indication of emotional intelligence. 

\textbf{I connected with the agent.} As reasons and indications for why they felt more connected with the affective agent, participants mentioned a notion of caring for the affective agent's reactions, having a visual connection with it, and its entertainment value nurturing this connection. One participant describes this in the following way: \emph{`` I actually wanted to have this (affective) one to look happy, so I didn't want it to look sad so that is definitely a kind of connection I would say'')} (P13) and another participant stated \emph{``Like a pet, it was very much a tamagotchi vibe. go faster and it was "hugh, I better go faster", I think that is somehow a connection even though that it is not a high level connection''} (P6). 

Overall, participants' explanations were in line with our observations that the affective agent required \textbf{more effort to interact} with, both in terms of \textbf{performing a voice act and understanding or interpreting the response behavior}. One participant described their takeaways in the following way: \emph{``It (affective agent) may make you feel like you always have to be on, switched on, to interact with it. It "requires" you to be emotionally engaged to interact with it in that it responds to your tone and emotions, which then makes you feel like you have to be on. It takes more efforts to interact with which could be a negative of it. But it just depends how your feeling and reciprocate''} (P3). These observations are also reflected in the participants' rating for feelings of control, understanding, and collaboration. Participants' attention was mainly captured by the affective agent with the standard agent being perceived as tagging along and robotic. This is also reflected in the following statement of a participant: \emph{``I guess this one felt more like a robot, did whatever it was programmed to do, of course this one did that too but it, because it reacted to the way I said it with emphasis it would move faster. I felt like it supported the way I was saying things''} (P12). The agent's richness in \textbf{expressions} can be described as \textbf{communicating a sort of social agency} which was intriguing for users and the playful nature of its expressions (which has been attributed to its behavior and definitely to the display of the emojis) was perceived as having an \textbf{entertaining value} and potentially as a \textbf{reward system}.      

\section{Discussion}

We argued in the beginning that speech command interfaces make little use of humans’ vocal range. However, as humans, we can combine what we say and how we say it to communicate efficiently and expressively with each other. Thus, our motivation to look at speech command and speech emotion as input modalities in combination is somewhat similar to Richard Bolt’s~\cite{Bolt_put_that_there_1980} seminal research on combining speech and gestures. In our context, speech emotions are like ``gestures'' which we have used to define ``How''  the affective agent moves, such as speed and ``how'' the agent looks (using emojis) in response to ``How'' the user talks to the agent. 

The body of research studying affective interactions with embodied agents and how such agents affect users is large (e.g.,~\cite{BEALE2009755}) and documents a plethora of variables, such as social bonding, perseverance, and caring in an education setting~\cite{Burleson, koschmann2017computer}. The field of studying speech emotions for affective interaction, especially outside the traditional dialog design is under explored. Our research is inspired by recent research~\cite{aslan2025speejis} addressing this gap in a human-human communication setting where speech emotions are used to augment and provide emotional context to human communication. We build on this previous research to study the use of speech emotions in a human-agent interaction setting where the user commands an agent.

To this end, this paper presents and evaluates examples of embodied agents that can recognize and respond to both speech command and speech emotions to better connect with users. Our initial motivation was to use the additional speech emotion modality to provide users  acknowledgment for the subjectivity in their expressions, rendering otherwise mundane experiences more reflective and alive, similar to how emojis can achieve this in human-human messaging. 

\subsection{User Experience}
The results of the UEQ questionnaire are in line with some of our expectations (e.g., significant increase in stimulation) based on what has been reported in previous research~\cite{aslan2025speejis}. We came to understand that it is a more challenging setting to use speech emotions to augment an agent's behavior (e.g., modify agent`s movement) compared to augmenting a humans message. There can be many reasons, including a potential bias towards machine intentions and concerns of users interacting with an artificial agent at the same social level, but also that this is a new situation in which users have to calibrate their affective interactions from scratch, requiring more effort and attention.      

Participants considered using emotional speech commands as harder to get familiar with and requiring more effort to learn to interact with an affective agent compared to interaction with the standard agent (i.e., Perspicuity). Further, in contrast to the standard agent, the affective agent's behavior was perceived as less `reliable' and `controllable' (i.e., Dependability) in terms of users being able to predict the exact behavior of the affective agent. However, it is somewhat surprising that the ratings are significant, considering that the affective agent implements all commands as successful as the standard agent (i.e., reaching the target) and enables the user to command the affective agent so that it can reach the target faster through increasing the arousal in their voice. The analysis of the post-hoc interviews shed more light on participants' reasons for their ratings, enabling us to infer relevant design considerations. It became clear that the frame users set for rating both agents were very different, as they connected with the agents in different ways, making it somewhat hard to compare the concrete ratings about the pragmatic and task-oriented use of these agents, which arguably was shadowed by the higher levels of engagement that the affective agent achieved. On the other hand, participants rated the interaction with the affective agent as more `fun', `exciting', and `motivating' (i.e., Stimulating), which demonstrates an added quality to a speech interface or product.

\subsection{Reflections on Attribution}
The affective agent copied or contained a part of the user by embodying the users' ``emotions''. One could argue that users perceive a ``skewed'' part of their own expressions in the behavior and feedback of the affective agent. The feelings associated with this experience were engaging and stimulating. While in our setup the stimulation and engagement was described as positive, there were a few occasions where it was perceived as something negative---for example, when a participant interpreted the affective agent's behavior as defiant. The use of speech emotions was perceived as fostering a strong connection, while the emotions and experiences created were described as entertaining. We believe this is due to the playful interaction setup, visualization, and the use of emojis. However, we have also become aware that there is potential to induce other kinds of strong experiences through the use of speech emotions, including negative sentiments such as those associated with creepiness or self-doubt, which could still be useful in a gaming setting, for example. 

As speech emotions and speech commands are intrinsically linked, it is impossible or difficult to provide a speech command without exposing speech emotions. It is possible to mask one's speech emotions with skill and deliberate voice acting, which is what participants felt they had to do in our study. With speech technologies being augmented with the ability to extract speech emotions, as humans, we may need to be more aware of when and how we expose such information to a machine and indirectly to the various stakeholders of such machines. In a co-located human-human setting, we are often aware of contextual factors, such as differentiating between work and leisure, but when we interact with machines or websites the context of the interaction is not as clear.   

 We saw that using speech emotions as part of speech commands asks users to be more careful and master their voice deliberately. Speech emotions create a connection between users and the agent. Arguably, this connection is enforced on both sides, asking users to invest in their interaction by using effortful voice acting and self-styling on the one side, and on the other, the system is enabled to respond with an emotionality that carries some form of intelligence familiar to the user.

\subsection{Design Considerations and Challenges}

Participants mentioned that the design invited them to explore and experiment. It was argued that this was potentially due to the interaction techniques novelty and its  stimulation of interest and curiosity. In this context, we observed~\textbf{serendipitous discoveries} of how participants figured out to gain more control over the affective agent, e.g. they would add the word ``please'', or stating words such as ``slower'' and ``faster'' in
embodied ways. While we told participants that the agent only understood a few keywords, they were still disappointed by the affective agent only when it reacted to the tone of their voice but did not implement the concrete command. For example, on one occasion a participant yelled ``stop'' and the affective agent reacted by listening to the arousal in the participant's voice and going faster. The behavior was perceived as less acceptable than doing it as the standard agent, which was perceived as kind of pretending that it did not hear the participant at all. 

Considering the qualitative analysis, we formulate the following design considerations for agents implementing the concept of emotional speech commands:

\begin{itemize}
\item The feeling of control is refined by higher expectations and more effortful involvement and attention of the users. Users will feel less control, which depending on the context is not necessarily a bad thing. Arguably the feeling of control has to be re-framed to a feeling and concept of ``joint control''~\cite{Seemann09} requiring an attitude of ``mutual trust'', which is arguably challenging to negotiate in a human-agent interaction setting.
\item Feelings of collaboration can increase due to the additional agency associated with the agent. Even if users feel less control, this makes them feel that the agent has more potential for collaboration. One participant explained this with the example that they don´t have control over other humans as well and still feel that they can collaborate with other humans. This consideration supports results from previous research~\cite{Melo11} emphasizing that emotions displays only can already be used to identify cooperators.
\item Feelings of engagement both positive and negative increase with such affective agents, even if the agent only uses non-verbal communication. Similar results are  also reported in related work where an agent’s co-speech gestures and their impact on user engagement and interest in the overall system have been studied~\cite{Aicher24}.
\item Feeling that the agent understands one depends on the congruence or incongruence of linguistic and acoustic expression. In the congruent case it can foster feelings of being understood on a higher level, similar to feelings of empathy, with the opportunities of designing ``empathic technologies'' and experiences~\cite{Jinan25}.  
\item Independent of the agent's concrete reaction, the agent will be perceived as  emotionally intelligent and as ``looking smart''~\cite{Haviland1976}, similar to what parents experience in their infants' facial expressions when they achieve a new task, such as grasping something. However, the agent might be perceived as less intelligent overall, as intelligence is seen as a complex construct. For example, when participants' expectations where purely task driven) the standard agent was perceived as more responsible  and competent, because it didn't react to emotional expressions and it didn't ``ask'' for attention.  
\item Feelings of connection are highly influenced. Especially when agents display signs of sadness user care and want to change that while when the agent is overly expressive it can be seen both as dramatic and entertaining. In any case the affective agent fosters connection. Arguably, the connection relates to the agent's impact on the user's attention, fostering mindfulness and reflection, which would be in line with research linking attention and social connection~\cite{Quaglia15} in human-human interactions.
\end{itemize}

While the design considerations are based on episodic interactions, we have to also evaluate the results considering the bigger picture in which conversational user interfaces are becoming ubiquitous and speech is an accessible modality in various contexts. What people say combined with how they say exposes people to machines and indirectly to all the stakeholders. Speech carries nuances in affect and allows one to track affect trajectories throughout contexts. Computer audition as a field is expanding with machines to perceive what is being said, how it is said and the scene in which it is said. On the positive side, these information can be used to provide users highly adaptive and situated service, help and assistance.

As with any affective computing design in which human emotions are extracted and utilized there is some risk and ethical issues involved. While the interaction space is being extended, people's ability and personality are different. A~\textbf{new level of accessibility} and related issues are introduced, resulting in people having a form of \textbf{performance anxiety} and doubting that they may not have the ability to perform the required way.  
We have also understood, based on our observations and discussion with our participants, that there is a conflict considering the perception of control, and questions \textbf{who is controlling who}. Some participants inherently disliked the idea of being controlled by having to say something in a specific emotional way and that this way of interacting is reserved for something else, than interacting with machines.

\subsection{Limitations}
Limitations of this research include the fact that we haven´t given our study participants any time to master their ``voice acting'' and did not introduce details of how the way they command would impact the agents behavior. Multiple participants mentioned that they struggled at first to understand the mapping between how they said something and the corresponding reactions of the affective agent. If we had given the participants more time to train before asking them to complete a more concrete task, the ratings may have been different. Our setup resulted in participants taking into account some initial frustration, but also the positive side of interacting with something novel and surprising. On the other hand, we were curious about how different users would manage using emotional speech commands in their own way.
Although the task was simple, it seemed appropriate for a `walk up and use' study, encouraging users to experiment and make serendipitous discoveries.
In retrospect, we could have implemented additional commands considering commanding movements, such as ``slower'', ``faster'', and ``stop'', so that participants did not have to experience these incongruent situations where what the participants stated was not in accordance with how they said it.
At this stage in our studies, we refrained from using LLMs. However, in the future, we expect to see more related research using LLMs, which can arguably provide easier access based on prompts to explore affective conversational user interfaces~\cite{Votintseva24}. Thus,  there is some potential to improve the linguistic connection between users and agents even for non-verbal responses, which represents important future research.

\section{Conclusion}
This paper explored the concept of emotional speech commands, comparing users' experiences in a simple task commanding two agents back and forth between two targets at opposite ends of a screen. In our user study the agents were embodied playfully as retro-style vehicles carrying an emoji.  Both the standard and the affective agent listened and implemented the command with the affective agent changing its ``mimic'' (i.e., emoji) and movement pattern in response to the speech emotion modality. We have explored and described the impact of expanding a speech command interface into an emotional speech command interface, where speech emotions are used not to overwrite the command, but to render the implementation of the command more social and entertaining.  To the best of our knowledge, this is the first study exploring the concept of emotional speech commands, and as such, the study has been more exploratory, providing initial results, but also opening new directions for discoveries. The prospect of profiling users' speech emotion transitions and delivering impactful user experiences and affective behavior requires a careful and responsible application. 


\bibliographystyle{ACM-Reference-Format}
\bibliography{references}


\end{document}